\documentclass[prl,graphicx,amsmath,amssymb,preprint]{revtex4-1}

\usepackage{graphicx}
\usepackage{color}

\draft % marks overfull lines with a black rule on the right

\begin{document}

%\preprint{}

\title{Non-relativistic tachyons: a new representation of the Galilei group} %Title of paper
%Tachyon representations are not so inconsistent,  



\author{V. Aldaya}
% \homepage{http://www.Second.institution.edu/~Charlie.Author.}
\email{valdaya@iaa.es}
\affiliation{%
Instituto de Astrof\'\i sica de Andaluc\'\i a (IAA-CSIC), Glorieta de la Astronom\'\i a, E-18080 Granada, Spain%\\This line break forced% with \\
}%
\author{J. Guerrero}%
 \email{julio.guerrero@ujaen.es}
\affiliation{ 
Departamento de Matem\'aticas, Universidad de Ja\'en, Campus las Lagunillas, 23071 Ja\'en, Spain%\\This line break forced with \textbackslash\textbackslash
}%
\altaffiliation[Also at ]{Institute Carlos I of Theoretical and Computational Physics, University of Granada, Fuentenueva s/n, 18071 Granada, Spain}

\author{F.F. L\'opez-Ruiz}
\email{paco.lopezruiz@uca.es}
 \affiliation{Departamento de F\'\i sica Aplicada, Universidad de C\'adiz, Campus de Puerto Real, E-11510 Puerto Real, C\'adiz, Spain}%Lines break automatically or can be forced with \\


\date{\today}

\begin{abstract}
An algebraic characterization of the contractions of the Poincar\'e group permits a proper construction of a non-relativistic limit of its tachyonic representation. We arrive at a consistent, nonstandard representation of the Galilei group which was disregarded long ago by supposedly unphysical properties. The corresponding quantum (and classical) theory shares with the relativistic one their fundamentals, and serves as a toy model to better comprehend the unusual behavior of the tachyonic representation. For instance, we see that evolution takes place in a spatial coordinate rather than time, as for relativistic tachyons, but the modulus of the three-momentum is the same for all Galilean observers, leading to a new dispersion relation for a Galilean system. Furthermore, the tachyonic objects described by the new representation cannot be regarded as localizable in the standard sense.

\end{abstract}

\pacs{Valid PACS appear here}% insert suggested PACS numbers in braces on next line

\maketitle %\maketitle must follow title, authors, abstract and \pacs

\section{Introduction}

Even though the concept of faster-than-light particles had been considered previously, the term `tachyon' was originally coined by Feinberg in 1967 \cite{Feinberg1}. That paper initiated a thorough discussion in the literature that continued over the seventies \cite{Aharonov,Sudarshan1,Sudashan2,Schroer,Jue,Feinberg2}, although it slowly faded away due to concerns on causality and, mainly, to the unsuccessful implementation of Lorentz transformations on the quantum theory of the tachyon scalar field. 

After that, tachyons have been arising every now and then in string theory \cite{Sen, Armoni}, supersymmetry \cite{ellis}, cosmology \cite{Paddy,Teixeira}, spontaneous symmetry breaking \cite{Felder1,Felder2}, QCD \cite{Yokota} and other areas \cite{Oriekhov,Trachenko}. There are even recent attempts to solve the failure in the representation of the Lorentz group on the Hilbert space of the tachyon field theory \cite{Dragan2}.

In two previous papers \cite{tachymom,prd}, we have shown explicit realizations of the tachyonic representations of the proper, orthochronous Poincar\'e group, $\mathcal{P}^\uparrow_+$, that is, representations associated with the one-sheeted hyperboloid as coadjoint orbit. These representations were classified long ago by Wigner in 1939, although he left their construction undone in his original paper \cite{Wigner}. Later authors did give a construction in momentum space (see, for instance, \cite{Fonda}), although the methods used were, in most cases, somewhat heuristic. In our paper \cite{tachymom}, such a representation was associated with an unconventional central extension $\tilde{\mathcal{P}}^\uparrow_+$ of $\mathcal{P}^\uparrow_+$ by $U(1)$. Being these representations constructed on the basis of an algebraic scheme, a Group Approach to Quantization \cite{23,2004,CMP0,CMP1,vallareport}, a suggestion of a quite different kinematics naturally arises which makes the tachyon theory less unphysical. The main difference lays in that the system evolves in a spatial direction and the time coordinate acquires character of a dynamical variable, canonically conjugated to $P_o$. Also a proper scalar product was previously introduced in Ref. \cite{prd} for the realization in the configuration space. 

In this paper, we aim at trying to understand the seemingly paradoxical idea of a non-relativistic limit of tachyons. To this end, we accomplish algebraically the limit of the tachyonic representations of the Poincar\'e group, $\mathcal{P}^\uparrow_+$, by means of an In\"on\"u-Wigner contraction \cite{inonuwigner} of the specific central extension $\tilde{\mathcal{P}}^\uparrow_+$.% with respect to the ordinary (algebraic) non-relativistic limit of the relativistic tachyons. 

We arrive, this way, at a new and unconventional (even exotic) class of representations of the Galilei group which was rarely discussed and, even less, constructed, to our knowledge. To be precise, in \cite{Levy-Leblond63} there is a mention to some sort of internal energy of the free Galilean particle which now could be related somehow to a contribution from the new representation (see below).  Besides that, the only reference that the authors found regarding the new representation of the Galilei group explicitly is Ref. \cite{inonuwignergalileo}, where In\"on\"u and Wigner found it but was never constructed. It was discarded as they considered that it does not contain particle states.

Apart from the intrinsic interest of explicitly computing this non-relativistic limit, it serves as a toy model to experience the essential rationale underlying tachyonic kinematics at a lower computational cost.

\section{The quantum representation for Galilean tachyons}

Let us start with the Lie algebra of a central extension $\tilde{\mathcal{P}}^\uparrow_+$ of the Poincar\'e group exhibiting the basic commutators among the generators of the representations corresponding to scalar tachyons \cite{tachymom}: 
\begin{align*}
	[P_{o},P_{i}] &=0, &[K_{i},P_{j}] &=\delta_{ij} P_{o} + \lambda_o \delta_{ij}\Xi, \\ 
	[P_{o},J_{i}] &=0, 
	&[K_i,P_o] &=P_{i} + \lambda_i \Xi\\
	[P_{i},P_{j}] &=0, 
	&[J_{i},P_{j}] &=\eta_{ij\cdot}^{\phantom{ij}k} P_{k} + \eta_{ij\cdot}^{\phantom{ij}k} \lambda_k \Xi \\
	[J_{i},K_{j}] &=\eta_{ij\cdot}^{\phantom{ij}k} K_{k}, 
	&[J_{i},J_{j}] &=\eta_{ij\cdot}^{\phantom{ij}k} J_{k} %+ \eta_{ij\cdot}^{\phantom{ij}k} n_k \Xi
	\\
	& &[K_{i},K_{j}] &=-\eta_{ij\cdot}^{\phantom{ij}k} J_{k} %- \eta_{ij\cdot}^{\phantom{ij}k} n_k \Xi
	\,,
\end{align*}
where $P_o$, $P_i$ stand for the generators of space-time translations, $K_i$ for boosts, $J_i$ for spatial rotations and $\Xi$ generates $U(1)$. We fix $\hbar=1$. %{\color{red}We set $\vec n=0$, which otherwise labels the intrinsic spin of the representation.}
 $\lambda_\mu\equiv(\lambda_o,\vec \lambda)$ is a four-vector in the co-algebra that will parametrize the different representations: $\lambda_\mu$ characterizes the three scalar co-adjoint orbits, that is, the one- or two-sheeted hyperboloid and the cone, according to the Lorentz character of $\lambda_\mu$. In order to select the scalar tachyonic representation at the Lie algebra level, we must consider space-like $\lambda_\mu$. For the sake of simplicity,  we shall fix $\lambda_o=0$ (see \cite{tachymom}).

 We are interested in a non-relativistic limit of the scalar tachyonic representation. In order to find it, we perform a specific In\"on\"u-Wigner contraction of the centrally extended Poincar\'e Lie algebra for tachyons, which proceeds by selecting the subgroup generated by $(P_o, \vec J, \Xi)$, multiplying by $\frac{1}{c}$ all the generators except those of the selected subgroup, and taking the limit $c\rightarrow\infty$. In intuitive terms, we can say that this group contraction is the standard non-relativistic limit (versus the Carrollian one, that will be discussed in a forthcoming work), making boosts \textit{and} spatial translations \textit{small} with respect to time translations, rotations and the quantum phase. The remaining extended algebra corresponds to the Lie algebra of a peculiar central extension of the Galilei group (actually a `pseudo' extension, see below), which will lead to a new representation:
\begin{align}
	[P_{o},P_{i}] &=0, &[K_{i},P_{j}] &=0, \nonumber\\ 
	[P_{o},J_{i}] &=0, 
	&[K_i,P_o] &=P_{i} + \lambda_i \Xi\nonumber\\
	[P_{i},P_{j}] &=0, 
	&[J_{i},P_{j}] &=\eta_{ij\cdot}^{\phantom{ij}k} P_{k} + \eta_{ij\cdot}^{\phantom{ij}k} \lambda_k \Xi \label{algebragalitaqui}\\
	[J_{i},K_{j}] &=\eta_{ij\cdot}^{\phantom{ij}k} K_{k}, 
	&[J_{i},J_{j}] &=\eta_{ij\cdot}^{\phantom{ij}k} J_{k} \nonumber\\
	& &[K_{i},K_{j}] &=0\,.\nonumber
\end{align}

We already observe at this level that $K_i$, $P_j$ are no longer canonically conjugated generators, as their commutator does not produce a central generator $\Xi$, whereas $\vec \lambda \wedge \vec J$, $\vec \lambda \wedge \vec P$, on the one hand, and $\vec \lambda \cdot \vec K$, $P_o$, on the other, do play this role since the commutator gives rise to a central term (here, `$\wedge$' indicates 3-vector product). In particular, $P_o$ turns out to acquire basic dynamical content, that is, character of canonically conjugated generator. In other words, the centrally extended Lie algebra obtained above points to a representation of the Galilei group in which the time variable is the coordinate of a dynamical degree of freedom, rather than an evolution parameter.

Although the actual construction of the new representation might be achieved in different ways, we choose to follow a Group Approach to Quantization \cite{23,2004,CMP0,CMP1,vallareport}, which starts from the group law of the relevant symmetry group. The Lie algebra \eqref{algebragalitaqui} can be exponentiated to a group law of the form of a central extension $\tilde G$ of $G \ni g $ by $U(1)\ni  \zeta $: 
\begin{align*}
	g''&=g' * g\,,   \\
	\zeta''&= \zeta' \zeta e^{i\xi(g',\,g)}\,, 
	\label{ley}
\end{align*}
where $g'*g$ corresponds to a standard group law for the Galilei group in terms of the parameters $t$, $\vec x$ for space-time translations, $\vec v$ for boosts and $\vec \epsilon$ for rotations \cite{23,2004,CMP0,CMP1,vallareport}, and the cocycle is actually a coboundary generated by $\delta(g)=i\vec \lambda\cdot \vec x$. 

We must remark that there exists a specific class of coboundaries (to be referred to as pseudo-cocyles) that plays a relevant role in the theory of group representations. In particular, in connection with the Poincar\'e group representations, these pseuco-cocycles were first analyzed by E.J. Saletan \cite{saletan}. Pseudo-cocycles were also required in order to unitarize some representations, for instance in conformal theories including Kac-Moody group representations \cite{mickelsson}. A complete algebraic analysis of these pseudo-cocyles in relation with the coadjoint orbits of the corresponding groups can be seen in Ref.~\cite{2004}.

Then, a direct computation leads to left- and right-invariant generators ($\mathbb I$ is the identity $3\times3$ matrix): 

\begin{align}
	\tilde X^L_{t}&= \frac{\!\!\partial}{\partial t}+ \vec v\cdot \frac{\!\!\partial}{\partial \vec x}+ \vec \lambda\cdot\vec v \;\Xi \nonumber
	&\tilde X^R_{t}&= \frac{\!\!\partial}{\partial t}	
\\
	\tilde X^L_{x^i}&= R(\vec \epsilon\,)^j_i\Big[\frac{\!\!\partial}{\partial x^j}+
	(\mathbb I-R(\vec \epsilon\,)^{-1})_j^k \lambda_k \Xi  \Big] \nonumber
		&\tilde X^R_{\vec x}&= \frac{\!\!\partial}{\partial \vec x} 
\\
	\tilde X^L_{v^i}&= R(\vec \epsilon\,)^j_i\frac{\!\!\partial}{\partial v^j}	
&\tilde X^R_{\vec v}&= \frac{\!\!\partial}{\partial \vec v}+t\frac{\!\!\partial}{\partial \vec x}
	+t\vec{\lambda}\;\Xi
\\
	\tilde X^L_{\vec \epsilon}&=  X^{L\,SU(2)}_{\vec \epsilon} \nonumber
	&\tilde X^R_{\vec \epsilon}&=  X^{R\,SU(2)}_{\vec \epsilon} 
	+ \vec x \wedge  \frac{\!\!\partial}{\partial \vec x}
	+ \vec v \wedge  \frac{\!\!\partial}{\partial \vec v}+
	\vec x \wedge \vec \lambda \Xi \nonumber
\\
	\tilde X^L_{\zeta}&= i\zeta \frac{\!\!\partial}{\partial \zeta}-i\zeta^*\frac{\!\!\partial}{\partial \zeta^*}\equiv \Xi \nonumber
&\tilde X^R_{\zeta}&= i\zeta \frac{\!\!\partial}{\partial \zeta}-i\zeta^*\frac{\!\!\partial}{\partial \zeta^*}\equiv \Xi \,. 
	\label{leftrightX}
\end{align}

Right-invariant generators may be chosen to act on complex $U(1)$-functions on $\tilde G$, $\Psi=\zeta \Phi (g)$, as derivations (as (pre-)quantum operators), leading to a reducible representation. In fact, given that on any Lie group all left-invariant generators commute with the right-invariant ones and act non-trivially on this representation, invoking Schur's  lemma the representation is reducible.  Then, we can choose a maximal, compatible left-subalgebra to be trivialized on the representation, that is to say, to impose a first-order Polarization which consistently reduces the representation. Such subalgebra actually exists: 
\[
\mathbb P^1=\langle \vec \lambda \cdot \tilde X^L_{\vec x}\,,\vec\lambda\cdot\tilde X^L_{\vec \epsilon}\,,\vec \lambda\wedge\tilde X^L_{\vec v}\,;\tilde X^L_{t}\,,\vec\lambda\wedge\tilde X^L_{\vec x}\rangle\,,
\]
constituted by the characteristic subalgebra (kernel of the Lie-algebra cocycle) $\mathcal C = \langle \vec \lambda \cdot \tilde X^L_{\vec x}\,,\vec\lambda\cdot\tilde X^L_{\vec \epsilon}\,,\vec \lambda\wedge\tilde X^L_{\vec v}\rangle$ and half of the canonically conjugated left-generators, divided into Moment-like $\langle\tilde X^L_{t}\,,\vec\lambda\wedge\tilde X^L_{\vec x}\rangle$ and Coordinate-like $\langle \vec \lambda \cdot \tilde X^L_{\vec v}\,,\vec\lambda\wedge\tilde X^L_{\vec \epsilon}\rangle$. 

After a bit of computations, the conditions $\tilde X^L \Psi =0$, $\tilde X^L \in \mathbb P^1$ reduce $\Psi = \zeta \Phi(g)$ to have the form
\[
\Phi = e^{-i(\mathbb I-R(\vec \epsilon\,) ) \vec \lambda \cdot \vec x} e^{-i R(\vec \epsilon\,) \vec \lambda \cdot \vec v t}\varphi(\vec \lambda\wedge R(\vec \epsilon\,)\vec \lambda\,, R(\vec \epsilon\,)\vec \lambda\cdot \vec v)\,,
\]
where $\varphi$ is an arbitrary function of the ``momentum'' variables
\[
\vec P_\perp\equiv \frac{\vec \lambda}{|\vec \lambda|}\wedge R(\vec \epsilon\,)\vec \lambda\,, \qquad 
P_t \equiv R(\vec \epsilon\,)\vec \lambda\cdot \vec v\,.
\]
The variable $P_t$ has the dimensions of an energy ($\vec \lambda$ has the dimensions of a momentum) although it is a basic quantity and not a derived function as in the standard Galilean representation. We also see that the variable $\vec P=R(\vec \epsilon\,)\vec \lambda$ (which lies on the surface of a sphere $S^2$ and therefore has only two independent components) is naturally decomposed in two parts: that perpendicular to $\vec \lambda$, $\vec P_\perp$, on which the reduced wave function arbitrarily depends, and $P_\parallel \equiv \vec \lambda \cdot R(\vec \epsilon\,)\vec \lambda$, which can be written in terms of $\vec P_\perp$. Hence, the support of the wave functions in the Hilbert space has the topology of a tree-dimensional cylinder $S^2\times \mathbb R$ (see later), replacing in the Galilean limit the one-sheet mass hyperboloid of the scalar tachyonic representation of the Poincar\'e group.

It should be pointed out that in Ref. \cite{Levy-Leblond63} an extra constant term in the energy of the free particle $\frac{1}{2} m v^2 + \mathcal V$ was considered as internal energy, although the corresponding representation is equivalent to the standard one. However, a Noether invariant associated with time translation would acquire  this form, $\frac{1}{2} m v^2 + P_t$, but with $P_t$ non-constant, in a central extension of the Galilei group associated with a cocycle sum of the ordinary one plus the present cocycle (coboundary, indeed). This situation would appear when considering the Galilean limit of the tachyonic representation of the Poincar\'e group characterized by space-like $\lambda_\mu$  with $\lambda_o\neq 0$. Such a Galilean representation is not equivalent to the one here considered, with $\lambda_o=0$, because Lorentz transformations do not commute with the Galilean limit.

The new irreducible representation is obtained by letting the right-invariant generators act on the reduced wave functions, with the following result for the basic ones: 
\begin{align}
\hat P_t\, \varphi &= i \tilde X^R_t \varphi=R(\vec \epsilon\,)\vec \lambda\cdot \vec v\, \varphi \equiv P_t \,\varphi\,, \nonumber
\\
\hat{\vec P}_\perp\,\varphi&=i \vec \lambda\wedge\big(\tilde X^R_{\vec x}-i \vec \lambda\big)\,\varphi = 	\vec \lambda\wedge R(\vec \epsilon\,)\vec \lambda\,\varphi \equiv \vec P_\perp \varphi\,,\nonumber
\\
\hat{\vec J}_\perp\,\varphi&=i \frac{\vec \lambda}{|\vec \lambda|}\wedge \tilde X^R_{\vec \epsilon}\,\varphi = \sqrt{\vec \lambda^2-\vec P^2_\perp}	\frac{\vec \lambda}{|\vec \lambda|}\wedge  \frac{\!\!\partial}{\partial \vec P_\perp}\,\varphi \,,
\\
\hat{K}_\parallel\,\varphi&=i \frac{\vec \lambda}{|\vec \lambda|}\cdot \tilde X^R_{\vec v}\,\varphi = \frac{\sqrt{\vec \lambda^2-\vec P^2_\perp}	}{|\vec \lambda|} \frac{\!\!\partial}{\partial  P_t}\,\varphi \,,\nonumber
\end{align}
where we have redefined the generator $\tilde X^R_{\vec x}$ in order to restore the extended $U(1)$-correction, and the position operators have been named so as to remind their origins in the Galilei group.  

The rest of the operators, those in the characteristic subalgebra, are non-basic and can be written in terms of the basic ones above: 
\[
\hat P_\parallel =i \vec \lambda \cdot \tilde X^R_{\vec x}\equiv i \tilde X^R_{x_\parallel}= \sqrt{\vec \lambda^2-\hat{\vec P}_\perp^2}\,, \quad 
\hat{\vec K}_\perp= \frac{\hat K_\parallel \hat{\vec P}_\perp}{\sqrt{\vec \lambda^2-\hat{\vec P}_\perp^2}}\,,\quad 
\hat J_\parallel= \frac{\hat{\vec J}_\perp \cdot\hat{\vec P}_\perp}{\sqrt{\vec \lambda^2-\hat{\vec P}_\perp^2}}\,,
\]
expressions which are well defined since the operators involved do commute. The function $\sqrt{\vec \lambda^2-\vec P^2_\perp}$ turns out to be the Noether invariant associated with the evolution generator $\vec \lambda \cdot \tilde X^R_{\vec x}\equiv \tilde X^R_{x_\parallel}$, here playing the role of ``Hamiltonian''. Note that various operators involve a square root and both signs are to be taken into account for their global definition. 

In order to achieve the corresponding, equivalent representation in configuration space, we have to seek a higher-order polarization \cite{polas}: 
\[
\mathbb P^{HO}\equiv
 \langle \tilde X^{L\,HO}_{x_\parallel}\,,\vec \lambda \cdot\tilde X^L_{\vec \epsilon}\,,\vec \lambda \wedge\tilde X^L_{\vec v}\,;
 \vec \lambda\cdot \tilde X^L_{\vec v}\,,\vec \lambda \wedge \tilde X^L_{\vec \epsilon}\rangle\,,
\]
which contains the generators of the undesired variables (momenta, in this case), but one of the generators in the characteristic subalgebra has been replaced by a second-order one: 
\[
\tilde X^{L\,HO}_{x_\parallel} \equiv (\vec \lambda\cdot  X^L_{\vec x})^2+(\vec \lambda \wedge X^L_{\vec x})^2+\vec\lambda^4\,,
\]
where for the sake of simplicity we have written the translation left-generators with the central extension contribution substracted ($ X^L_{\vec x}$ without tilde). 

The polarization conditions coming from the first-order generators in $\mathbb P^{HO}$ simply tell us that $\Phi(g)$ does not depend on $\vec v$ nor $\vec \epsilon$, then remaining the higher-order condition, which leads to
\begin{equation}
\tilde X^{L\,HO}_{x_\parallel} \Psi =0 \quad \Rightarrow \quad \big(\nabla^2_{\vec x}+\vec \lambda^2\big)\phi(t,\vec x)=0
\label{Helmholtz_eq}
\end{equation}
once we have restored the `rest translation momentum'  as $\Phi= e^{-i \vec \lambda\cdot\vec x}\phi(t,\vec x)$ (analogous to the rest-energy restoration $\Phi= e^{i \lambda_o x_o}\phi(t,\vec x) \equiv e^{i m c^2 t}\phi(t,\vec x)$ of the traditional tardyonic representation). This equation, the Helmholtz equation in 3D, plays the the same role as that of the time-dependent Schr\"odinger equation for the standard quantum Galilean free particle.  However, \eqref{Helmholtz_eq} should be interpreted as the evolution equation in the variable $x_\parallel$ without any restriction to the dependence on $t$ beyond the condition of square-integrability of $\phi$ once established the scalar product (see below). The Helmholtz equation for a quantum system had already appeared in the description of a free particle on a sphere in the momentum space representation \cite{S3mom} (see \cite{symmetry} for a comparison of the Helmholtz equation with the Klein-Gordon equation for both for $m^2>0$ and $m^2<0$). It has also appeared recently in the context of Carrollian \cite{FOFCarroll} and Galilean fields \cite{FOF}.

The action of the right-invariant generators on $\phi$ becomes: 
\begin{align}
\hat P_t\, \phi &= i \frac{\partial\phi}{\partial t} \,, \nonumber
\\
\hat{\vec P}_\perp\,\phi&= \vec \lambda\wedge i \frac{\partial\phi}{\partial \vec x}\,,\nonumber
\\
\hat{\vec J}_\perp\,\phi&=\frac{\vec \lambda}{|\vec \lambda|}\wedge \big(\vec x \wedge i \frac{\partial\phi}{\partial \vec x}\big) \,,
\\
\hat{K}_\parallel\,\phi&= t \vec \lambda \cdot i \frac{\partial\phi}{\partial \vec x} \,.\nonumber
\end{align}

The other operators, as in the case of the momentum representation, are functions of the basic ones. 

It should be noticed that this representation of the Galilei group inherits from the relativistic one, among other features, the necessity of a non-local scalar product \cite{S3mom,prd,WolfLibro}. The reason for that lays again in the absence of some range of momenta: the modulus of $\vec P_\perp$ is bounded by $\lambda\equiv |\vec \lambda|$ and only `limited-band' modes are reachable within the Hilbert space. This is also related to the fact that it is not possible to achieve a Dirac delta in configuration space.

In fact, the standard, and natural, scalar product on momentum space, adapted to the present case, becomes

\begin{equation}
\langle \varphi, \varphi' \rangle = \int_{\mathbb R^4} d^4P \delta \big(P_\parallel^2 + {\vec P}_\perp^2 - \lambda^2 \big) \varphi^* \varphi' = 
	\int_{\Omega_\lambda} \frac{d P_t d^2 P_\perp}{2 |P_\parallel|} \big(\varphi^*_+ \varphi'_+ + \varphi^*_- \varphi'_- \big)\,,
\label{PescMom}
\end{equation}
where $P_\parallel = \pm \sqrt{\lambda^2-{\vec P}_\perp^2}$, $\Omega_\lambda$ is a  solid  cylinder $D_{|\vec \lambda|}\times \mathbb R$ (where $D_{|\vec \lambda|}$ indicates a disk of radius $|\vec \lambda|$), and $\varphi_\pm=\varphi(\pm |P_\parallel|,P_t,\vec P_\perp)$. As we have mentioned, the three-momentum $\vec P$ is restricted to lay on a sphere of radius $\lambda$, so that the modulus of the transverse momentum $\vec P_\perp \in  D_{|\vec \lambda|}$ and is therefore bounded from above. In the Galilean limit, these restrictions do not affect the ``energy'' $P_t$ which, we insist, is now an independent variable and therefore must appear in the scalar product as an integration variable.

In going to the configuration space, the Fourier transform in the $P_t$ variable behaves in the direct ordinary way, since the integration measure does not depend on $P_t$ and the range of $P_t$ is unconstrained. However, in the $\vec P_\perp$ plane, the situation is quite different due to the restriction of $\vec P$ to a two-dimensional sphere (see \cite{S3mom,WolfLibro}), and some Bessel functions appear as integration kernels in a non-local form of scalar product. We get for $\phi(t,\vec x_\perp, x_\parallel)$ and $\phi'(t',{\vec x}\,'_\perp, x_\parallel)$ the following: 

\begin{equation}
\begin{split}
		\langle \phi,\phi'\rangle = 
			C \int d^2x_\perp d^2x'_\perp dt\, dt'
			&\Big( \phi^*(t,\vec x_\perp, x_\parallel) \frac{J_{3/2}(\lambda |\vec x_\perp-\vec x{\,}_\perp'|)}{|\vec x_\perp-\vec x{\,}_\perp'|^{3/2}} \delta(t-t') \phi'(t',{\vec x}\,'_\perp, x_\parallel) 
			\\ 
	+  &\dot\phi^*(t,\vec x_\perp, x_\parallel)  \frac{J_{1/2}(\lambda |\vec x_\perp-\vec x{\,}_\perp'|)}{\lambda|\vec x_\perp-\vec x{\,}'_\perp|^{1/2}} \delta(t-t')\dot \phi'(t',{\vec x}\,'_\perp, x_\parallel) \Big)
	\\
	= 
	C \int d^2x_\perp d^2x'_\perp dt 
	&\Big( \phi^*(t,\vec x_\perp, x_\parallel) \frac{J_{3/2}(\lambda |\vec x_\perp-\vec x{\,}_\perp'|)}{|\vec x_\perp-\vec x{\,}_\perp'|^{3/2}} \phi'(t,{\vec x}\,'_\perp, x_\parallel) 
	\\ 
	 & +\dot\phi^*(t,\vec x_\perp, x_\parallel)  \frac{J_{1/2}(\lambda |\vec x_\perp-\vec x{\,}_\perp'|)}{\lambda|\vec x_\perp-\vec x{\,}'_\perp|^{1/2}} \dot \phi'(t,{\vec x}\,'_\perp, x_\parallel) \Big)\,,
	\label{PescConfig}
\end{split}	
\end{equation}
where the dot over the wave functions $\phi$ means derivative of $\phi$ with respect to $x_\parallel$, which can be computed on the surface $x_\parallel=0$ since this product does not depend on $x_\parallel$, and the constant $C$ can be fixed in order to make the Fourier transform between momentum and configuration space unitary, giving $C=\sqrt{\frac{2\pi^3}{\lambda}}$. 

%{\color{red}Nocion de probabilidad de presencia debe ser revisada,  interpretacion probabilistica?}

Several comments are in order. First, in close analogy with the results in Refs. \cite{S3mom,prd}, the Fourier transform unitarily mapping the Hilbert space in momentum representation into the one in configuration representation is given by
\begin{align}
\label{FT}
\phi(t,\vec x_\perp) &= \frac{1}{(2\pi)^{3/2}}\int d P_t \int_{D_{|\vec \lambda|}}\frac{d^2 P_\perp}{2 |P_\parallel|} 
	\Big(\varphi_+(P_t,\vec P_\perp)+\varphi_-(P_t,\vec P_\perp)\Big)e^{i\vec P_\perp \cdot \vec x_\perp}e^{-i P_t t}\;,
	\\
	\dot \phi(t,\vec x_\perp) &= \frac{-i}{(2\pi)^{3/2}}\int d P_t \int_{D_{|\vec \lambda|}}\frac{d^2 P_\perp}{2} 
	\Big(\varphi_+(P_t,\vec P_\perp)-\varphi_-(P_t,\vec P_\perp)\Big)e^{i\vec P_\perp \cdot \vec x_\perp}e^{-i P_t t},
\label{FTd}
\end{align}
where $\phi(t,\vec x_\perp) $ and $\dot\phi(t,\vec x_\perp) $ are $\phi(t,\vec x_\perp, x_\parallel)$ and its derivative with respect to $x_\parallel$ evaluated at $x_\parallel=0$. The corresponding inverse transform can be written
\begin{equation}
\label{IFT}
	\varphi_\pm(P_t,\vec P_\perp)=\!\frac{1}{(2\pi)^{3/2}}\!\!
	\int \!\!dt \int d^2x_\perp \Big(\! |P_\parallel|_+ \phi(t,\vec x_\perp)\pm 
	i |P_\parallel|_+^0 \dot \phi(t,\vec x_\perp)\!\Big)e^{-i\vec P_\perp \cdot \vec x_\perp}e^{i P_t t}\,,
\end{equation}
where $|P_\parallel|_+ = |P_\parallel|$ for $|\vec P_\perp\,|\leq \lambda$, $|P_\parallel|_+ = 0$ for $|\vec P_\perp\,|\geq \lambda$, $|P_\parallel|_+^0 = 1$ for $|\vec P_\perp\,|\leq \lambda$ and $|P_\parallel|_+^0 = 0$ for $|\vec P_\perp\,|\geq \lambda$. %Eqs. \eqref{IFT}, \eqref{FT} and \eqref{FTd} define a `Fourier' transform and its inverse, relating configuration and momentum spaces. In particular, it can be checked that eigenstates of momentum $\vec p\,'$ in configuration space are mapped into eigenstates of momentum $\vec p\,'$ in momentum space and vice versa. 

Second, only the set of oscillatory solutions of \eqref{Helmholtz_eq}, equipped with the scalar product \eqref{PescConfig}, can be mapped unitarily through \eqref{FT}, \eqref{FTd} and \eqref{IFT} into the set of functions on the solid cylinder which are integrable with respect to the scalar product in momentum space \eqref{PescMom}. By oscillatory solutions we mean those whose Fourier spectrum is limited to real $P_\parallel=\pm \sqrt{\lambda^2-{\vec P}_\perp^2}$. Also, the scalar product \eqref{PescConfig} can be shown to be positive definite for oscillatory solutions.

\bigskip

\section{Classical viewpoint}

Before deepening in the quantum theory, let us gain some insight about the classical limit of this Galilean tachyonic system, which also serves to the particular aim of the present paper, that is, of being a toy model for thinking of tachyons. The GAQ provides the classical theory in the midway to quantization. In fact, the quantization form $\Theta$ (see \cite{23,2004,CMP0,CMP1,vallareport}) is a generalization of the Poincar\'e-Cartan 1-form of the involved physical system. By taking the quotient by those generators in the characteristic subgroup (the kernel of $\Theta$ and $d\Theta$) other than the proper evolution, we obtain the Poincar\'e-Cartan (classical) form $\Theta_{PC}$ up to a total differential: 
\begin{equation}
	\label{tetapc}
	\Theta_{PC}= \vec p_\perp \cdot d\vec x_\perp - p_t dt + \sqrt{\lambda^2-\vec p\,^2_\perp} d x_\parallel\,,
\end{equation}
where we can read the `Hamiltonian'  $H= - \sqrt{\lambda^2-\vec p\,^2_\perp}$ determining the dynamical evolution in $x_\parallel$ and the degrees of freedom. 

This form is invariant under the generators $Y_a$ (right-invariant generators $\tilde X_a^R$ with the $\zeta$-component suppressed and after having taken the quotient by the above-mentioned subgroup), $L_{Y_a} \Theta = 0$: 

\begin{align}
	Y_{t}&= \frac{\!\!\partial}{\partial t}\,,	 &\nonumber
	Y_{\vec \epsilon}&=  X^{R\,SU(2)}_{\vec \epsilon} 
	+ \vec x \wedge  \frac{\!\!\partial}{\partial \vec x}
	+ \vec v \wedge  \frac{\!\!\partial}{\partial \vec v}+
	\vec x \wedge \vec \lambda \Xi \,,\nonumber
\\
	Y_{\vec x_\perp}&= \frac{\!\!\partial}{\partial \vec x_\perp} \,, &
	Y_{\vec p_\perp}&= \vec p_\perp\frac{\!\!\partial}{\partial p_t}+t\frac{\!\!\partial}{\partial \vec x_\perp}\,, \label{leftrightX}
\\
	Y_{x_\parallel}&= \frac{\!\!\partial}{\partial  x_\parallel}\,, & \nonumber
	Y_{p_\parallel}&= \sqrt{\lambda^2-\vec p\,^2_\perp}\frac{\!\!\partial}{\partial p_t}+t\frac{\!\!\partial}{\partial  x_\parallel}\,,
\\
	Y_{\zeta}&= i\zeta \frac{\!\!\partial}{\partial \zeta}-i\zeta^*\frac{\!\!\partial}{\partial \zeta^*}\equiv \Xi \,, \nonumber
\end{align}
which obviously close the Lie algebra of the unextended Galilei group. 

In any case, we may adopt \eqref{tetapc} and \eqref{leftrightX}, from scratch, as the characterization of the simplest classical system realizing the tachyonic Galilean mechanics. The classical trajectories are the integral curves of the field $X$ so that $i_X d\Theta_{PC}=0$, with $X$ being
\[
X=\frac{\!\!\partial}{\partial  x_\parallel}+\frac{\vec p_\perp}{\sqrt{\lambda^2-\vec p\,^2_\perp}}\cdot \frac{\!\!\partial}{\partial \vec x_\perp}\,,
\]
giving the equations of motion
\begin{align}
\frac{d \vec x_\perp}{dx_\parallel}&=\frac{\vec p_\perp}{\sqrt{\lambda^2-\vec p\,^2_\perp}}\,, & \frac{dt}{dx_\parallel}&=0\,,
\\
\frac{d \vec p_\perp}{dx_\parallel}&=0\,, & \frac{d p_\parallel}{d x_\parallel}&=0\,.
\end{align}

Note that the rules of Mechanics apply in a standard way except for the fact that the evolution parameter is not $t$ but a particular  spatial coordinate $x_\parallel=\vec\lambda\cdot \vec x$, arbitrarily chosen by the direction of $\vec \lambda$. Note also that, should we consider the possibility of describing this system in the context of a standard time evolution in $t$, we would be led to something like $\frac{dx_\parallel}{dt}=\infty$. This is consistent with an infinite standard velocity as the natural  non-relativistic limit of a faster-than-light speed. However, we must be aware once again of the status of $t$ in this context: $t$ is a coordinate, not an evolution parameter, which suggests the notion of tachyons being objects whose trajectories are spread over space rather than time: much in the same way the trajectory of a standard particle is an extended object in time $t$ (we are referring to the world line of the particle), the trajectory of a tachyonic mechanical system should be an extended object in the spatial dimension $x_\parallel$ and, in general, $\vec x$. More on this when returning to the quantum theory.

If one prefers the Lagrangian description, the change of variables
\[
\dot{\vec x}_\perp\equiv \frac{\partial H}{\partial\vec p_\perp}= \frac{\vec p_\perp}{\sqrt{\lambda^2-\vec p\,^2_\perp}}\,; \quad \dot t \equiv 	\frac{\partial H}{\partial p_t} = 0
\]
 leads to the Lagrangian (dot means derivative with respect to $x_\parallel$): 
 \[
 L = \vec p_\perp \cdot \dot{\vec x}_\perp + \sqrt{\lambda^2-\vec p\,^2_\perp} = \lambda\sqrt{1+\dot{\vec x}\,^2_\perp}\,, 
\] 
which has a manifest formal resemblance with the known Lagrangian in the action for the free tardyonic relativistic particle $ m_o \int d\tau =  m_o\int \sqrt{c^2dt^2-d\vec x\,^2}=  m_o c \int \sqrt{1-\frac{\vec v\,^2}{c^2}} dt$. In fact, if we consider a `proper space' interval $dr$, we can see that it plays an analogous role to proper time $\tau$:  $\int dr=\int\sqrt{d\vec x\,^2}=\int \sqrt{1+\dot{\vec x}\,^2_\perp}dx_\parallel $, and we can say that the trajectories for Galilean tachyons minimize the `proper space' interval. 

The Noether invariants can be computed by $i_{Y_a}\Theta_{PC}$: 
\begin{align}
	\vec P_\perp &\equiv i_{Y_{\vec x_\perp}}\Theta_{PC} = \vec p_\perp\,, & \nonumber
	P_\parallel &\equiv i_{Y_{x_\parallel}}\Theta_{PC} = \sqrt{\lambda^2-\vec p\,^2_\perp}=-H\,,\\
	\vec K_\perp &\equiv i_{Y_{\vec p_\perp}}\Theta_{PC} = t\; \vec p_\perp\,, &
	K_\parallel &\equiv  i_{Y_{p_\parallel}}\Theta_{PC} =t \, \sqrt{\lambda^2-\vec p\,^2_\perp}\,,\\
	J_\parallel &\equiv  i_{Y_{\epsilon_\parallel}}\Theta_{PC} = \vec \lambda\cdot(\vec x_\perp\times \vec p_\perp)\,,&
	\vec J_\perp &\equiv  i_{Y_{\vec\epsilon_\perp}}\Theta_{PC} = x_\parallel \vec p_\perp-\vec x_\perp \sqrt{\lambda^2-\vec p\,^2_\perp}	\,,\nonumber\\
	P_t &\equiv - i_{Y_t}\Theta_{PC}=p_t \,.\nonumber
\end{align}

The relationship among Noether invariants already shows a peculiar property of the Galilean tachyonic objects: $|\vec P| = \lambda$, so they must have constant modulus of the three-momentum for all inertial reference frames, in close analogy with the constancy of $c$ in Special Relativity. We also see that $\vec K=t\vec P$. 

 The reason behind of $|\vec P|$ being constant is that the generators of the Galilean boosts,  $Y_{\vec p_\perp}$ and $Y_{p_\parallel}$, do not change $\vec p$. This can be made more explicit by computing the action of a finite Galilean transformation of parameters $B$ (time translation), $\vec A$ (spatial translation), $\vec V$ (boosts) and $\vec \epsilon$ (rotations $R(\vec \epsilon\,)$) acting on the classical tachyonic system: 
\begin{align}
	t'&= t + B \,,\nonumber\\ 
	\vec x\,' &= R(\vec \epsilon\,) \vec x + \vec V t + \vec A\,,\label{finite}\\
	{p_t}'&= p_t + R(\vec \epsilon\,) \vec p \cdot \vec V\,, \nonumber\\
	\vec p\,' &= R(\vec \epsilon\,) \vec p\,. \nonumber
\end{align}

The first two lines in \eqref{finite} are the usual ones, although we must be aware that the evolution parameter $x_\parallel$ is now affected by spatial rotations and boosts together with spatial translations. In the last line, we see that the transformation for $\vec p_\perp$ and $H$ is compatible with the preservation of $|\vec p\,|$ under boosts. Finally, the transformation of $p_t$ can be compared to the standard transformation law for the energy in Galilean particles of mass $m$: $E'=E + R(\vec \epsilon\,) \vec p \cdot \vec V+ \frac{1}{2}m \vec V^2$.

Another comment  is in order: \eqref{finite} is in perfect agreement with the coadjoint action on a coadjoint orbit of the Galilei Lie group which has been constructed very recently \cite{FOF} (see Sec. 7.2.2 with $k=0$). This puts the new representation introduced in this paper on an equal footing with the well-known representations associated with its central extensions of mass $m$ for standard Galilean particle.

As far as the negativity of the energy is concerned, we must be aware, once again, of the role played by $P_t$:  $P_t$ is unbounded from below, but it is just a time momentum. On the contrary, the quantity $P_\parallel$, the actual Hamiltonian, keeps a proper bounding both from below and from above. Thus, this system differs from a usual mechanical system (with unbounded from above energy).

\section{Quantum states}

Returning to the quantum description of Galilean tachyons, we can make some considerations on the dispersion relation involved, given by

\[
P_\parallel^2 + \vec P^2_\perp = \vec P^2=\lambda^2, \qquad P_t \; \hbox{independent of $\vec P$}\,, 
\]
which is nothing but the consequence of the Helmholtz equation \eqref{Helmholtz_eq}. First, given the action of the Galilei group on momenta \eqref{finite}, the dispersion relation is equivalent for all Galilean observers since it is preserved by any Galilean transformation. Second, $P_t$ being independent of $\vec P$ means that any quantum state given by $\phi (t,\vec x)$ factorizes into
\[
\phi(t,\vec x) = f(t) \psi(\vec x)\,,
\]
where $f(t)\in L^2 (\mathbb R)$ and $\psi(\vec x)$ is an oscillatory solution of the Helmholtz equation normalizable with respect to the scalar product \eqref{PescConfig}. Therefore, a change in the `frequency' ($P_t$) spectrum of a state does not necessarily imply a different spatial pattern of the wave function. %In other words, should a frequency-fixed excitation be introduced in the system, it would only produce modes of zero momentum $\vec P$ and infinite wavelength.
This should be compared with the standard representation of the Galilei group, where the Schr\"odinger equation implies $E=\frac{\vec p\,^2}{2m}$, and therefore the frequency spectrum ($E$) and the spatial spectrum ($\vec p\,$) are related. 

\bigskip

We now give some relevant solutions of the Helmholtz equation \eqref{Helmholtz_eq}
$\psi(\vec x)$. The most localized wave function in momentum space for (say) positive $P_\parallel=+\sqrt{\lambda^2-{\vec P}_\perp^2}$ is given by
\[
\varphi_+^{pw+}(P_t,\vec P_\perp)= 2P_\parallel \delta^{(2)}(\vec P_\perp-\vec P_\perp')\delta(P_t-P_t')\,, \qquad
\varphi_-^{pw+}(P_t,\vec P_\perp)= 0\,,
\]
which is a distribution (non-integrable in the standard sense), and $\vec P_\perp'$ must satisfy $|\vec P_\perp'|<\lambda$. Note that if $P_\parallel=-\sqrt{\lambda^2-{\vec P}_\perp^2}$, we would have $\varphi_+^{pw-}(P_t,\vec P_\perp)= 0$ and $\varphi_-^{pw-}(P_t,\vec P_\perp)=\varphi_+^{pw+}(P_t,\vec P_\perp)$. Making use of the Fourier transform \eqref{FT} and \eqref{FTd}, we get
\[
\phi_{pw+}(\vec x_\perp,t)=\frac{1}{(2\pi)^{3/2}} e^{i\vec P'_\perp \cdot \vec x_\perp}e^{-i P_t t}\,, 
\qquad 
\dot \phi_{pw+}(\vec x_\perp,t)=\frac{-iP_\parallel}{(2\pi)^{3/2}}e^{i\vec P'_\perp \cdot \vec x_\perp}e^{-i P_t t}\,,
\]
which are nothing other than initial conditions at $x_\parallel=0$ for plane waves. Including the evolution in $x_\parallel$, we have an expression formally equivalent to the standard one:
\[
\phi_{pw+}(\vec x,t)=\frac{1}{(2\pi)^{3/2}} 
e^{i(\vec P'_\perp \cdot \vec x_\perp+\sqrt{\lambda^2-\vec P_\perp^{'2}} \,x_\parallel)}
e^{-i P_t t}
=\frac{1}{(2\pi)^{3/2}} e^{i\vec P' \cdot \vec x}e^{-i P_t t}\,.
\]

However, given that $P_t$ is independent of $\vec P$, we could have a superposition of spatially identical plane waves with different dependencies on $t$, to have the form $\frac{1}{2\pi} e^{i\vec P' \cdot \vec x}f(t)$, with $f(t)$ any square-integrable function, as pointed out above. Then, we can focus on just the spatial dependence: 
\[
\psi_{pw+}(\vec x) = \frac{1}{2\pi} e^{i\vec P' \cdot \vec x}\,.
\]
\bigskip

\begin{figure}[h!]
 \centering
   \includegraphics[width=12cm]{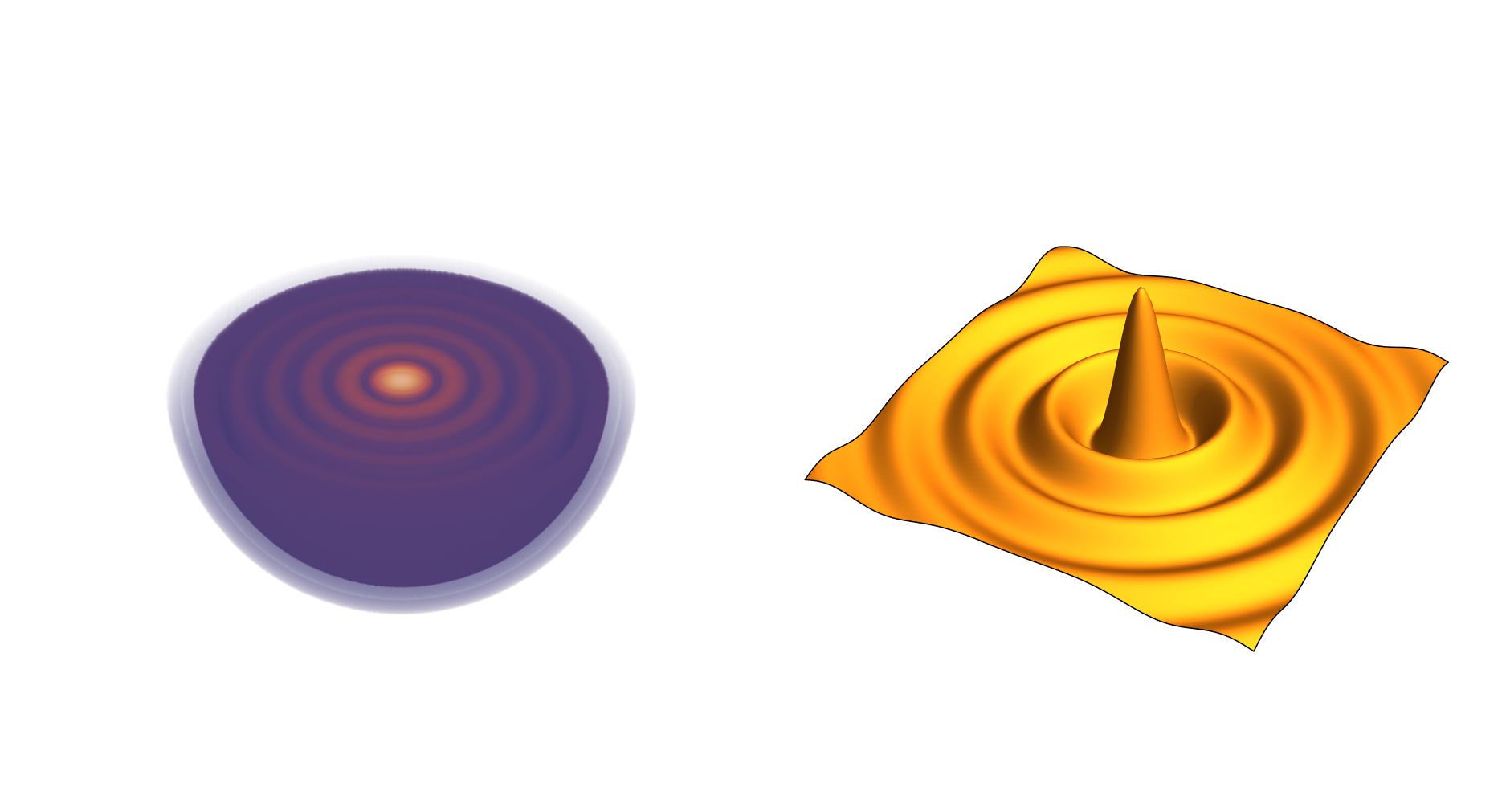}
   \caption{Depiction of the configuration space wave function of the most localized state $\psi_{loc}(\vec x)$, which is spherically symmetric,  using a cut of a density plot (left) and a 3D plot of a two-dimensional slice (right).  }
   \label{fig1}
 \end{figure}

After that, let us introduce an interesting solution of \eqref{Helmholtz_eq}, which is the most spread state in momentum space: a uniform function over $\vec P_\perp$, 
\[
\varphi_+^{loc}(\vec P_\perp)= \frac{1}{\sqrt{2\pi \lambda}}\,, \qquad
\varphi_-^{loc}(\vec P_\perp)= \frac{1}{\sqrt{2\pi \lambda}}\,,
\] 
where we now omit the dependence on $P_t$. This is the most localized possible state around the origin in configuration space. However, it is a normalizable function rather than a Dirac delta function. Again, making use of the Fourier transform \eqref{FT} and \eqref{FTd}, we get (omitting the Fourier transform in $t$ since it gives an arbitrary function of $t$ factor)
\[
\phi_{loc}(\vec x_\perp)= \frac{1}{2}\frac{J_{1/2}(\lambda |\vec x_\perp|)}{\sqrt{ |\vec x_\perp|}}\,, \qquad \dot \phi_{loc}(\vec x_\perp)=0\,.
\]
Those are initial conditions at $x_\parallel=0$ for an oscillatory solution of \eqref{Helmholtz_eq}. The corresponding $x_\parallel$-dependent state is given by
\[
 \psi_{loc}(\vec x)=\frac{1}{2}\frac{J_{1/2}(\lambda r)}{\sqrt{r}}=\frac{\sqrt{\lambda}}{\sqrt{2 \pi}} \hbox{sinc}(\lambda r)\,, 
\]
where $r=|\vec x|=\sqrt{x^2+y^2+z^2}=\sqrt{\vec x\,^2_\perp+x_\parallel^2}$, containing the evolution parameter $x_\parallel$. Indeed this is not a Dirac delta function but an oscillatory wave function, due to the lack of part of the spectrum in momenta. This most-localized wave function has asymptotic behavior $\psi \sim \frac{1}{r}$ for $r\to \infty$, so that we can say that they are spread over space. However, wave functions \emph{can} be localized in $t$ as $P_t$ ranges all over $\mathbb R$ and the scalar product is the standard for $L^2(\mathbb R)$ in coordinate $t$.  See Fig.~\ref{fig1}. 

\bigskip

\begin{figure}[h!]
 \centering
   \includegraphics[width=12cm]{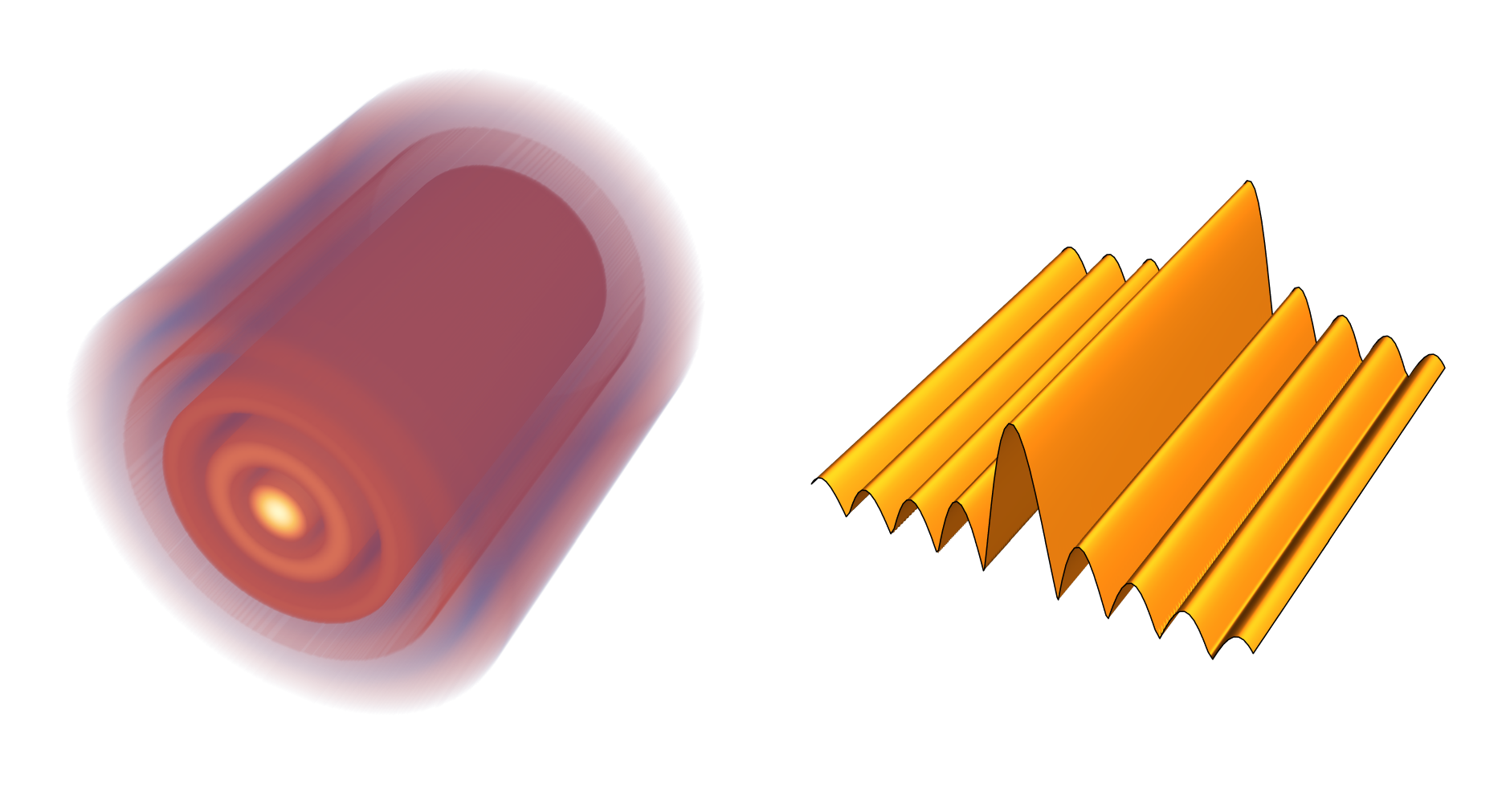}
   \caption{Depiction in configuration space of the Bessel wave $|\psi^{BW}_0(\vec x)|$  using a  a density plot (left) and a 3D plot of a two-dimensional slice (right).  }
   \label{fig2}
 \end{figure}

\begin{figure}[h!]
 \centering
   \includegraphics[width=12cm]{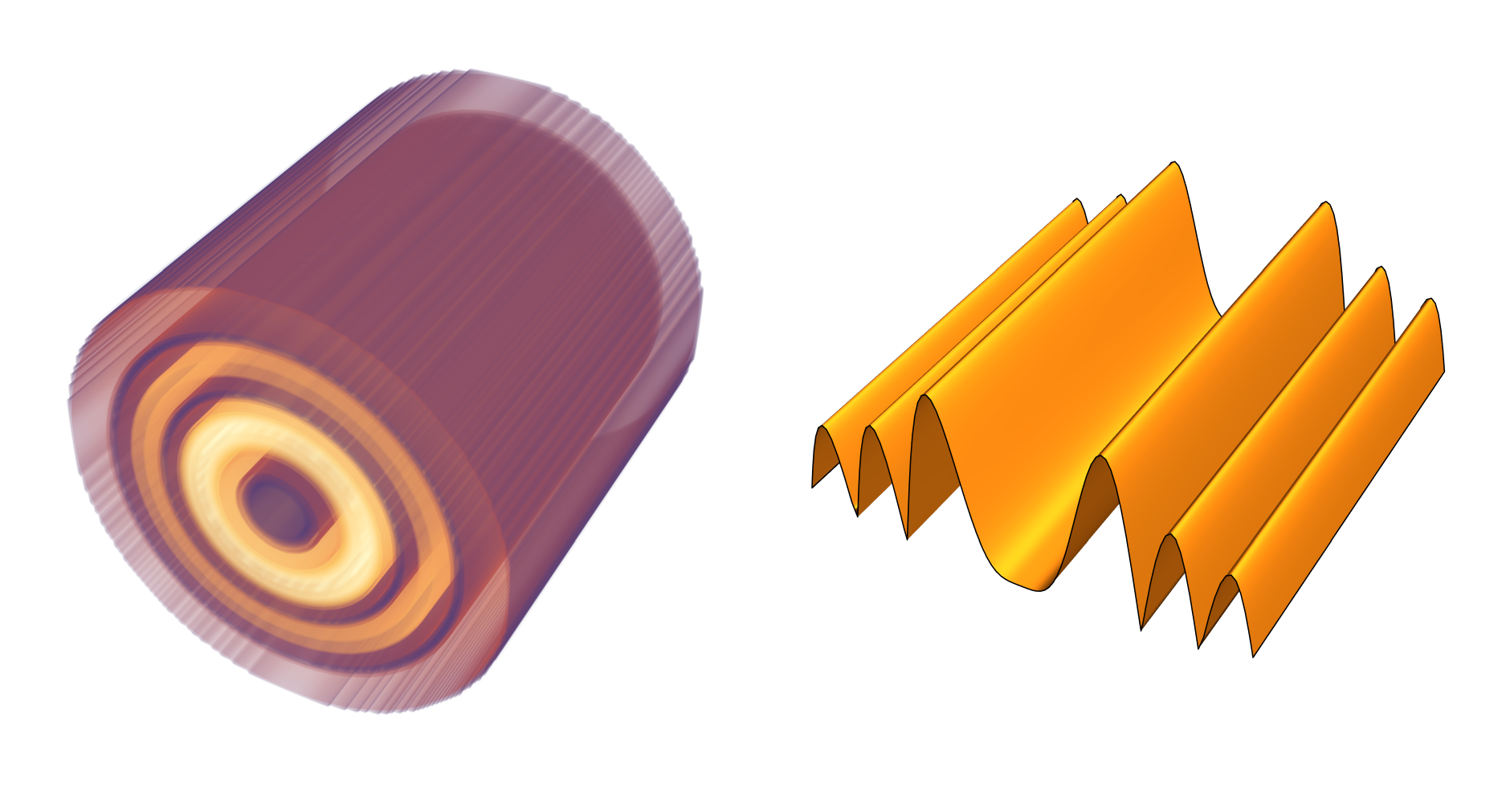}
   \caption{Depiction in configuration space of the Bessel wave $|\psi^{BW}_4(\vec x)|$  using a  a density plot (left) and a 3D plot of a two-dimensional slice (right).  }
   \label{fig3}
 \end{figure}

 Another peculiar state is given by Bessel beams \cite{opticajulio} oriented in an arbitrary direction (which does not necessarily coincides with $x_\parallel$), for instance $z$. Going directly to the spatial part of the wave function:  
 
 \[
 \psi^{BW}_m(\vec x)=J_{m}(\beta_{xy} \rho) e^{i m \theta} e^{-i\beta_z z}\,, 
 \]
 where $\theta=\arctan \frac{y}{x}$ is the polar angle in the plane $xy$, $\rho=\sqrt{x^2+y^2}$ and $\beta_{xy}^2+\beta_z^2=\lambda^2$. This state for a Galilean tachyon is a solution of \eqref{Helmholtz_eq} and is $x_\parallel$-dependent, where $x_\parallel$ can be identified with the coordinate along any Cartesian direction. $\psi^{BW}_m$ is fully spread in the direction $z$ although maximally localized in the transverse plane $xy$, and could be interpreted as a collimated beam. See Fig.~\ref{fig2} for $m=0$ and Fig.~\ref{fig3} for $m=4$.  This family of wave functions are not square integrable; rather, they are distributions, analogously to the plane waves described above. 

All solutions to the Helmholtz equation described here (and some others) have long been known in the literature in the context of Optics (see \cite{opticajulio} and references therein). However, the physical interpretation is quite different: these wave functions are required to be normalizable with respect to the scalar product \eqref{PescConfig} (or, at least, normalizable in the distributional sense), while in Optics the condition required is that of finite energy. Also, in the case of Galilean tachyons, the evolution parameter is not $t$: it is a spatial direction, and $t$ is just a coordinate. 

\section{Final remarks}

We have introduced a non-relativistic Galilean system which is a proper limit of the `$m^2<0$' representation of the Poincar\'e group, analyzing both its classical and quantum descriptions.
Although the actual physical model is the relativistic tachyonic system, the corresponding Galilean limit may serve as a simpler approximation showing tachyons described by a Galilean observer. There exists a different non-relativistic limit of the Poincar\'e group representations, namely the Carrollian limit \cite{Levy-Leblond65}, which will be studied elsewhere.
  
  Both relativistic and non-relativistic Galilean tachyons share some of the key features: in order to set up the variational problem, the time coordinate $t$ can not be taken as the evolution parameter and is replaced by an arbitrary spatial direction coordinate $x_\parallel$; also, the most spatially localized quantum states are not Dirac delta functions, they are oscillatory solutions of the Helmholtz equation instead, which now plays the role of the Schr\"odinger equation; they also share that a non-local scalar product is required in order to make the representation in configuration space unitary. 
  
  The non-relativistic tachyons here presented distinguish from the relativistic ones in that the quantum dispersion relation only involves spatial momenta $\vec P$, decoupled from $P_t$: the one-sheet mass hyperboloid goes to a three-dimensional cylinder in the Galilean limit. This can be compared with the non-relativistic limit of the two-sheet mass hyperboloid, which is the familiar paraboloid of the relation between kinetic energy and momentum in the free Galilean particle.

\bigskip

V.A. thank the Spanish Ministerio de Ciencia e Innovaci\'on (MICINN) for financial support (PID2022-116567GB-C22). J.G. and FF.L-R. acknowledge financial support from the Spanish MICINN (PID2022-138144NB-100).  V.A. also acknowledges financial support from the State Agency for Research of the Spanish MCIU through the `Center of Excellence Severo Ochoa' award for the Instituto de Astrof\'\i sica de Andaluc\'\i a (SEV-2017-0709).

% If in two-column mode, this environment will change to single-column format so that long equations can be displayed. 
%\begin{widetext}
%$$\mbox{put long equation here}$$
%\end{widetext}

% Figures should be put into the text as floats. 
% Use the graphics or graphicx packages (distributed with LaTeX2e).
% Put the label that you will use with \ref{} command in the braces of the \label{} command.
%
% \begin{figure}
% \end{figure}

% \begin{table}
% \end{tabular}
% \end{table}

%\begin{acknowledgments}
% Put your acknowledgments here.
%\end{acknowledgments}

%\bibliography{your-bib-file}

\begin{thebibliography}{11}


\bibitem{Feinberg1} G. Feinberg, \textit{Phys. Rev.} \textbf{159}, 1089 (1967).
\bibitem{Aharonov} Y. Aharonov, A. Komar and L. Susskind, \textit{Phys. Rev.} \textbf{182}, 1400 (1969).
\bibitem{Sudarshan1} M.E. Arons and E.C.G. Sudarshan, \textit{Phys. Rev.} \textbf{173}, 1622 (1968).
\bibitem{Sudashan2} J. Dhar and E.C.G. Sudarshan, \textit{Phys. Rev.} \textbf{174}, 1808 (1968).
\bibitem{Schroer} B. Schroer, \textit{Phys. Rev. D} \textbf{3}, 1764 (1971).
\bibitem{Jue} C. Jue, \textit{Phys. Rev. D} \textbf{8}, 1757 (1973).
\bibitem{Feinberg2}  G. Feinberg, \textit{Phys. Rev. D} \textbf{17}, 1651 (1978).
\bibitem{Sen} A. Sen, J. High Energy Phys. 08 (1998) 012. 
\bibitem{Armoni} A. Armoni and E. Lopez, \textit{Nuc. Phys. B} \textbf{632} (2002) 240.
\bibitem{ellis} J. Ellis et al., \textit{Phys. Rev. D} \textbf{78}, 075006 (2008).
\bibitem{Paddy} J.S. Bagla, H.K. Jassal and T. Padmanabhan, \textit{Phys. Rev. D} \textbf{67}, 063504 (2003).
\bibitem{Teixeira} E.M. Teixeira, A. Nunes and N.J. Nunes,  \textit{Phys. Rev. D} \textbf{100}, 043539 (2019).
\bibitem{Felder1} G. Felder et al., \textit{Phys. Rev. Lett.} \textbf{87}, 011601 (2001).
\bibitem{Felder2} G. Felder, L. Kofman and A. Linde, \textit{Phys. Rev. D} \textbf{64}, 123517 (2001).
\bibitem{Yokota} T. Yokota, T. Kunihiro and K. Morita, \textit{Phys. Rev. D} \textbf{96}, 074028 (2017).
\bibitem{Oriekhov} D.O. Oriekhov and L.S. Levitov, \textit{Phys. Rev. B} \textbf{101}, 245136 (2020).
\bibitem{Trachenko} C. Yang, M.T. Dove, V.V. Brazhkin and K. Trachenko,  \textit{Phys. Rev. Lett.} \textbf{118}, 215502 (2017).
\bibitem{Dragan2} J. Paczos, K. D\c{e}bski, S. Cedrowski, S. Charzy\'nski,K. Turzy\'nski,  A. Ekert and A. Dragan, \textit{Phys. Rev. D} \textbf{110}, 015006 (2024).
\bibitem{tachymom} V. Aldaya, J. Guerrero and F.F. L\'opez-Ruiz, \textit{Tachyons in ``momentum-space'' representation}, 	arXiv:2312.16522. 
\bibitem{prd} F.F. L\'opez-Ruiz, J. Guerrero and V. Aldaya, \textit{Phys. Rev. D} \textbf{102}, 125010 (2020). 
\bibitem{Wigner} E.P. Wigner, \textit{Ann. Math.} \textbf{40}, 149 (1939).  
\bibitem{Fonda} L. Fonda and G.C. Ghirardi, \textit{Symmetry Principles in Quantum Physics} (Marcel Dekker, New York, 1970), Chap. 5.
\bibitem{23} V. Aldaya and J.A. de Azcarraga, \textit{J. Math. Phys.} \textbf{23}, 1297 (1982).
\bibitem{2004} J. Guerrero, J.L. Jaramillo, and V. Aldaya,  J. Math. Phys. \textbf{45}(5), 2051-2072  (2004).
\bibitem{CMP0} V. Aldaya, J. Navarro-Salas and A. Ram\'\i rez, \textit{Commun. Math. Phys.} \textbf{121}, 541 (1989).
\bibitem{CMP1} V. Aldaya, M. Calixto and J. Guerrero, \textit{Commun. Math. Phys.} \textbf{178}, 399-424 (1996).
\bibitem{vallareport} V. Aldaya, M. Calixto, J. Guerrero and F.F. L\'opez-Ruiz, \textit{Int. J. Geom. Meth. Mod. Phys.} \textbf{8}, 1329-1354 (2011).
\bibitem{saletan} E.J. Saletan, \textit{J. Math. Phys.} \textbf{2}, 1 (1961).
\bibitem{mickelsson} J. Mickelsson, \textit{Phys. Rev. Lett.} \textbf{55}, 2099 (1985).
\bibitem{inonuwigner} E. In\"on\"u and E.P. Wigner, \textit{Proceedings of the National Academy of Sciences} \textbf{39}(6), 510 (1953). 
\bibitem{Levy-Leblond63} J.-M. L\'evy-Leblond, \textit{J. Math. Phys.} \textbf{4}(6), 776 (1963). 
\bibitem{inonuwignergalileo} E. In\"on\"u and E.P. Wigner, \textit{Il Nuovo Cimento} \textbf{9}, 705 (1952). 
\bibitem{polas} J. Guerrero and V. Aldaya, \textit{J. Math. Phys.} \textbf{41}, 6747 (2000).
%\bibitem{Mukunda} N. Mukunda, \textit{Ann. Phys.} \textbf{61}, 329 (1970).
%\bibitem{Landulfo} A.G.S. Landulfo, W.C.C. Lima, G.E.A. Matsas and D.A.T. Vanzella, \textit{Phys. Rev. D} \textbf{86}, 104025 (2012).
%\bibitem{Dragan1} A. Dragan, K. D\c{e}bski, S. Charzy\'nski, K. Turzy\'nski and A. Ekert,  \textit{Class. Quantum Grav.}, \textbf{40}(2), 025013 (2023).
%\bibitem{oscilata} V. Aldaya, J. Bisquert and J. Navarro-Salas, Phys. Lett. A, \textbf{156} (7-8), 381-385 (1991).
\bibitem{S3mom} J. Guerrero, FF. L\'opez-Ruiz and V. Aldaya, \textit{J. Phys. A} \textbf{53}, 145301 (2020). 
\bibitem{symmetry} FF. L\'opez-Ruiz, J. Guerrero and V. Aldaya,  \textit{Symmetry} \textbf{13}, 1302 (2021).
\bibitem{FOFCarroll} J. Figueroa-O'Farrill, A. Pérez and S. Prohazka, \textit{JHEP} \textbf{10}, 041 (2023).
\bibitem{FOF} J. Figueroa-O'Farrill, S. Pekar, A. Pérez and S. Prohazka, \textit{Galilei particles revisited}, arXiv:2401.03241 [hep-th]. 
\bibitem{WolfLibro} K.B. Wolf, \textit{Elements of Euclidean optics Lie Methods in Optics}, Lecture Notes in Physics, Heidelberg: Springer (1989), p 115.
\bibitem{opticajulio} B.M. Rodr\'\i guez-Lara, R. El-Ganainy and  J. Guerrero, \textit{Science Bulletin} \textbf{63}, 244–-251 (2018).
\bibitem{Levy-Leblond65} J.-M. L\'evy-Leblond, \textit{Ann. Inst. Henri Poincar\'e} \textbf{3}, 1 (1965). 

%\bibitem{S3config} V. Aldaya, J. Guerrero, FF. L\'opez-Ruiz and F. Coss\'\i o, \textit{J. Phys. A} \textbf{49}, 505201 (2016).
%\bibitem{2000} M. Calixto, V. Aldaya and M. Navarro, Int. J. Mod. Phys. A, \textbf{15}(25), 4011-4044 (2000).
%\bibitem{AnnPhys} V. Aldaya and J. de Azc\'arraga, Ann. Phys. \textbf{165}(2), 484-504 (1985).
%\bibitem{Moses} H.E. Moses, \textit{J. Math. Phys.} \textbf{9}, 2039 (1968).
%\bibitem{A26} V. Aldaya, J. Bisquert, J. Guerrero and J. Navarro-Salas, J. Phys. A \textbf{26}(20), 5375 (1993).
%%\bibitem{Wolf} S. Steinberg and K.B. Wolf, \textit{J. Math. Phys.} \textbf{22}, 1660 (1981). 
%%\bibitem{Mostafazadeh} A. Mostafazadeh and F. Zamani, \textit{Ann. Phys.} \textbf{321}, 2183 (2006).
%\bibitem{pethybridge} D. Anninos. T. Anous, B. Pethybridge and G. \c Seng\"or, \textit{J. Phys. A} \textbf{57}, 025401 (2024). 
%\bibitem{IJMP} M, Calixto, V. Aldaya and M. Navarro, \textit{Int. J. Mod. Phys. A} \textbf{15}(25), 4011-4044 (2000).
%\bibitem{EurPhys} V. Aldaya, \textit{Eur. Phys. J. Plus} \textbf{136}, 304 (2021). 
%\bibitem{GQ1} J.M. Souriau, \textit{Structure des systemes dynamiques}, Dounod, Paris (1970).  
%\bibitem{GQ2} D.J. Simms and N.M.J. Woodhouse, \textit{Lectures on Geometric Quantization}, Lecture Notes in Phys. vol. 52, Springer N.Y.
%\bibitem{GQ3} A. Pressly and G. Segal, \textit{Loop Groups}, Clarendon Press, Oxford (1986).
%\bibitem{Cartan} P. Malliavin, \textit{Géometrie différentielle intrinsique}, Hermann, Paris (1972).
\end{thebibliography}

\end{document}